\documentstyle[twocolumn,aps,prd]{revtex}
\begin{document}
\newcommand{\eq}{\begin{eqnarray}}
\newcommand{\en}{\end{eqnarray}}

\wideabs{
\title{On the Lifetime of the $\pi^+\pi^-$ Atom}

\author{A.\ Gall$^1$, J.\ Gasser$^1$, V.\ E.\ Lyubovitskij$^{2,3}$ and
A.\  Rusetsky$^{1,2,4}$}
\address{$^1$Institute for Theoretical Physics, University of Bern,
Sidlerstrasse 5, CH-3012, Bern, Switzerland}
\address{$^2$Bogoliubov Laboratory of Theoretical Physics, Joint Institute
for Nuclear Research, 141980 Dubna, Russia}
\address{$^3$Department of Physics, Tomsk State University, 
634050 Tomsk, Russia}
\address{$^4$HEPI, Tbilisi State University, 380086 Tbilisi, Georgia}

\date{\today}

\maketitle

\begin{abstract}
The $\pi^+\pi^-$ atom decays - in the ground state - predominantly into two 
neutral pions. We present a general  expression for the corresponding decay 
width in the framework of QCD (including  photons). It contains all terms 
at leading and next-to-leading order in isospin breaking. The result allows 
one to evaluate the combination $|a_0-a_2|$ of $\pi\pi$  $S$-wave scattering 
lengths  from   $\pi^+\pi^-$ lifetime measurements, like the one  presently  
performed by the DIRAC experiment at CERN.
\end{abstract}
\pacs{PACS number(s): 03.65.Ge, 03.65.Nk, 12.39.Fe, 13.40.Ks}
}

The DIRAC experiment at CERN~\cite{DIRAC} aims to measure the lifetime of the 
$\pi^+\pi^-$ atom in its ground state with high  precision. The atom decays 
predominantly into two neutral pions, 
$\Gamma=\Gamma_{2\pi^0}+\Gamma_{2\gamma}+\ldots,\;$ 
with $\Gamma_{2\pi^0}\simeq\Gamma$. The measurement will therefore 
allow one \cite{Deser,Uretsky} to determine the difference $|a_0-a_2|$ 
of the strong $S$-wave $\pi\pi$ scattering lengths with isospin $I=0,2$.
The value predicted \cite{ChPT} for this quantity may then be confronted with 
the one extracted from the lifetime measurement. What makes this enterprise  
particularly exciting is the fact that one may eventually determine in this 
manner  the nature of spontaneous chiral symmetry breaking in QCD by 
experiment: Should it turn out that the predictions  \cite{ChPT} are in 
conflict with the results of DIRAC, one will have to conclude that spontaneous
chiral symmetry breaking in QCD differs from the standard 
picture~\cite{Stern}.

In order to perform this test, the theoretical
expression for the  width must of course be known with a precision 
that matches 
the accuracy of  the lifetime measurement of DIRAC.
 It is the aim of this letter to fill existing gaps in this respect,
 and to derive a compact expression for $\Gamma_{2\pi^0}$ in the framework of
 QCD (including photons)  by use of effective field theory techniques.
 Our result contains all terms at leading and
 next-to-leading order in the isospin breaking parameters 
$\alpha\simeq 1/137$ and $(m_u-m_d)^2$.
 On the other hand, we expect that the contributions from 
 next-to-next-to-leading order 
 are completely negligible even for future improved data,
 and we  therefore discard them here.
 
Before describing our calculation, we briefly review previous work on
the subject. Theoretical investigations of hadronic atoms and, in particular, 
of $\pi^+\pi^-$ decays, have been performed in several settings. Potential 
scattering theory in the framework of quantum mechanics has  been used 
in~\cite{Deser,Trueman,Rasche,Minkowski}, and methods of quantum field theory 
have been invoked as well 
\cite{Efimov,Pervushin,Volkov,Jentschura,Sazdjian,Atom}.      
In particular, in Ref.~\cite{Sazdjian}, the lifetime of the 
$\pi^+\pi^-$ atom was calculated by use of two-body wave 
equations of 3D-constraint theory. In Ref.~\cite{Atom}, the $\pi^+\pi^-$ atom
decay was studied in a field-theoretical approach based on the 
Bethe-Salpeter equation. The results for the $\pi^+\pi^-$ atom lifetime
obtained with the two latter approaches contain the major
next-to-leading order terms in isospin breaking and agree both conceptually 
and numerically. However, the  investigations were restricted to the so called
"local" approximation, in which the momentum dependence of the strong
$\pi\pi$ scattering amplitude is neglected. The justification for this 
approximation is the huge difference between the momentum scale characterizing
the $\pi^+\pi^-$ system bound by the static Coulomb force ($\sim 1~{\rm MeV}$)
and the typical strong interaction scale of several hundred ${\rm MeV}$ that 
leads to a factorization \cite{Deser} in the observables of this type of bound
systems. The main difference between the results of the potential model 
approach in~\cite{Rasche} and  of the field-theory methods used 
in~\cite{Sazdjian,Atom} concerns  the correction due to the mass difference of 
the charged and neutral pions, see the discussion in~\cite{Atom}.
It remains to be seen whether the  discrepancy stems from an incomplete 
treatment of the isospin-breaking effects in the phenomenological strong 
potential used in~\cite{Rasche}.

In several recent publications~\cite{Labelle,Kong,Holstein}, the decay of 
$\pi^+\pi^-$ atoms has been  studied in the framework of a non-relativistic
effective Lagrangian - a method originally proposed by Caswell and 
Lepage~\cite{Lepage} to investigate bound states in general. This method has 
proven to be far more efficient for the treatment of loosely bound 
systems - such as the $\pi^+\pi^-$ atom - than conventional approaches 
based on relativistic bound-state equations. It allows one e.g. to go beyond
the local approximation used in~\cite{Sazdjian,Atom}. 
On the other hand, we are not aware of a systematic investigation of
the decay of the $\pi^+\pi^-$ atom in this framework. In particular, 
the chiral expansion of the width has not been discussed, and
a  comparison of the corrections found in this framework 
with the results of~\cite{Sazdjian,Atom} has never been provided.
Our letter intends to fill this gap.

In the following, we consider  QCD (including photons) in the two-flavor 
case. In this framework, the  width is a function of the fine-structure 
constant, of the quark masses and of the renormalization group invariant 
scale of QCD. Chiral perturbation theory \cite{ChPTlit} (ChPT) allows one to 
determine $\Gamma$ through a systematic expansion in powers of $\alpha$ and 
of the quark masses $m_u$ and $m_d$ (up to logarithms).
 As has been shown 
in Refs. \cite{Deser,Uretsky}, the leading term in $\Gamma_{2\pi^0}$ involves 
the square of $|a_0-a_2|$. We therefore write 
\eq\label{deser}
\Gamma_{2\pi^0}=\frac{2}{9}\,\alpha^3
p^\star{\cal A}^2(1+K),
\en
where $p^\star=(M_{\pi^+}^2-M_{\pi^0}^2-
\frac{1}{4}M_{\pi^+}^2\alpha^2)^{1/2}$, and where
the quantities ${\cal A}$ and $K$ are expanded in powers of $\alpha$ and 
$m_d-m_u$. In the following, we count $\alpha$ and $(m_d-m_u)^2$ as small 
parameters of order $\delta$. Then, the expansion for ${\cal A}$ and $K$ 
takes the form 
\eq\label{eq1}
{\cal A}&=&a_0-a_2+h_1\,(m_d-m_u)^2+h_2\,\alpha + o(\delta),
\nonumber\\[2mm]
K&=&f_1\,(m_d-m_u)^2+f_2\,\alpha\ln\alpha+f_3\,\alpha + o(\delta),
\en
where the scattering lengths $a_0$, $a_2$ and the coefficients $h_i$ and $f_i$ 
are evaluated at $m_u=m_d=\hat m$ and 
$M_\pi=M_{\pi^+} = 139.57$ MeV. The quantities $h_i$ and $f_i$
parameterize the corrections to the leading-order formula derived in 
\cite{Deser,Uretsky}. Below  we demonstrate that, for the calculation of the 
coefficients $f_i,h_i$ at all orders in the chiral expansion, it suffices to 
calculate the on-mass-shell scattering amplitude for the physical process
$\pi^+\pi^-\rightarrow \pi^0\pi^0$ in ChPT at order $\delta$. All coefficients
can then be unambiguously related to this amplitude, to  the $\pi^+-\pi^-$ 
mass difference, and to the strong scattering lengths $a_0$ and $a_2$. 

We  work in the Coulomb gauge and  eliminate the Coulomb photons by use of 
the equation of motion. The transverse photons, attached to the leading-order
diagrams of the strong interaction, do not contribute at the accuracy we are
interested here. They are therefore completely omitted in the following.
(This procedure is based on   a set of rules built on top of non-relativistic 
quantum field theory:  in the dimensional regularization scheme, the so called 
"threshold expansion"~\cite{Beneke} is applied to the diagrams, in the 
calculation of both, the scattering amplitudes and  the bound-state 
characteristics. This procedure guarantees the validity of power-counting 
rules.) With these provisions, the description of the $\pi^+\pi^-$ atom in the 
non-relativistic framework becomes remarkably simple - it is a purely 
two-channel problem with  charged and neutral pion fields. 

We proceed as follows. First, we display the non-relativistic effective pion 
Lagrangian, as derived from ChPT. Next, we formulate the resonance two-channel
$\pi\pi$ scattering theory by applying Feshbach's projection 
technique \cite{Feshbach}. This method allows one to explicitly display the 
pole structure of the scattering matrix element and to obtain the equation 
for the bound-state energy of the $\pi^+\pi^-$ atom. Solving this equation, 
the decay width of the atom is given in terms of the couplings in the 
non-relativistic Lagrangian. Finally, the width $\Gamma_{2\pi^0}$ is 
expressed in terms of the relativistic $\pi\pi$ scattering amplitude through 
the matching procedure.   
   
The non-relativistic effective Lagrangian 
${\cal L}={\cal L}_0 + {\cal L}_D + {\cal L}_C + {\cal L}_S$ - at the order 
of accuracy we are working here - consists of the free Lagrangian for  
charged and neutral pions (${\cal L}_0$), the "disconnected" piece 
(${\cal L}_D$) - providing  the correct relativistic relation between the 
energies and momenta of the pions - the Coulomb interaction piece 
(${\cal L}_C$), and the "connected" piece (${\cal L}_S$) which contains  
local four-pion interaction vertices:
\eq\label{Lagr_full}
{\cal L}_0&=& \sum_{i=\pm, 0}\,\pi_i^\dagger\biggl( i\partial_t-M_{\pi_i}+
\frac{\triangle}{2M_{\pi_i}}\biggr)\pi_i,\nonumber\\
{\cal L}_D&=&\sum_{i=\pm, 0}\,\pi_i^\dagger
\biggl(\frac{\triangle^2}{8M_{\pi_i}^3}+\cdots\biggr)\pi_i,\\
{\cal
L}_C&=&-4\pi\alpha(\pi_-^\dagger\pi_-)\triangle^{-1}(\pi_+^\dagger\pi_+)
+\cdots ,\nonumber\\
{\cal L}_S&=&c_1\pi_+^\dagger\pi_-^\dagger\pi_+\pi_-
+c_2[\pi_+^\dagger\pi_-^\dagger(\pi_0)^2+{\rm h.c.}]
+c_3\,(\pi_0^\dagger\pi_0)^2\nonumber\\
&+&c_4[\pi_+^\dagger\stackrel{\leftrightarrow}
{\triangle}\pi_-^\dagger(\pi_0)^2+\pi_+^\dagger\pi_-^\dagger\pi_0
\stackrel{\leftrightarrow}{\triangle}\pi_0+{\rm h.c.}]+\cdots,\nonumber
\en 
where $u\stackrel{\leftrightarrow}{\triangle}v\equiv 
u \triangle v + v \triangle u$. The coupling constants $c_i$ 
 are real at $O(\alpha)$ and are 
determined through matching to the relativistic theory. 
For illustration, we give the result of the tree-level
matching for $c_i$, including  isospin symmetry 
breaking corrections at order $p^2$ in ChPT:
\eq\label{matching}
& &c_1=\frac{1}{2F^2}(1+\kappa)+\cdots, \hspace*{.1cm} 
c_2=-\frac{3}{8F^2}\biggl(1+\frac{\kappa}{6}\biggr)+\cdots,\\
& &c_3=\frac{1}{16F^2}+\cdots,
\hspace*{1cm}
c_4=\frac{1}{32F^2M_{\pi^0}^2}(1-2\kappa)+\cdots,\nonumber
\en
where $F$ is the pion decay constant in the chiral limit, and 
$\kappa=M_{\pi^+}^2/M_{\pi^0}^2-1$. 
If the final result for the width is expressed
by use of Eq. (\ref{matching}), one has to keep in mind that the 
isospin breaking piece of e.g. $c_2$ contributes to the factor ${\cal A}$  
at order $\delta$ and must therefore be retained.

We now formulate the two-channel scattering theory. We denote the
full Hamiltonian derived from (\ref{Lagr_full}) by $H=H_0+H_C+V$, with
$V=H_D+H_S$. The scattering operator $T$  obeys the Lippmann-Schwinger  
equation $T(z)=(H_C+V)+(H_C+V)G_0(z)T(z)$. The free and the Coulomb Green 
operators are defined as $G_0(z)=(z-H_0)^{-1}$ and $G(z)=(z-H_0-H_C)^{-1}$, 
respectively. The pole structure of the $T$-matrix is predominantly 
determined by the static Coulomb interaction $H_C$, whereas $V$ generates 
a small shift of the pole positions into the complex $z$-plane and will be 
treated perturbatively. To this end, we  use the method developed by 
Feshbach~\cite{Feshbach} a long time ago. The $T$-matrix in our theory 
describes the transitions between  charged $|{\bf P},{\bf p}\rangle_+=
a^\dagger_+({\bf p}_1)a^\dagger_-({\bf p}_2)|0\rangle$ and neutral
$|{\bf P},{\bf p}\rangle_0=
a^\dagger_0({\bf p}_1)a^\dagger_0({\bf p}_2)|0\rangle$
states, where $a^\dagger_i$ denote the creation operators for 
non-relativistic pions. Further,   ${\bf P}={\bf p}_1+{\bf p}_2$
and ${\bf p}=\frac{1}{2}({\bf p}_1-{\bf p}_2)$ are the CM 
and relative momenta of pion pairs, respectively. We work in the
CM system and remove the CM momentum  
from the matrix elements of any operator $R$, introducing the notation 
\eq\label{cm}
_A\langle{\bf P},{\bf q}|R(z)|{\bf 0},{\bf p}\rangle_B=
(2\pi)^3\delta^3({\bf P})({\bf q}|r_{AB}(z)|{\bf p}),
\en
where $A,~B=+,~0$. The operators $r_{AB}(z)$ act in the Hilbert space
of vectors $|{\bf p})$, where  the scalar product is defined as the integral
over the relative three-momenta of pion pairs. 

In order to avoid the complications associated with charged particles
in the final states, we consider the elastic  scattering process 
$\pi^0\pi^0\rightarrow \pi^0\pi^0$. In the vicinity of the $\pi^+\pi^-$
threshold, the scattering  matrix element develops a pole at \cite{Feshbach}
\eq\label{main}
z-E_0-(\Psi_0|\tau_{++}(z)|\Psi_0)=0,
\en
where $({\bf p}|\Psi_0)=\Psi_0({\bf p})$ stands for the unperturbed Coulomb 
ground-state wave function, and  $E_0$ is the corresponding ground-state 
energy. According to the conventional definition, the decay width is  
$\Gamma=-2{\rm Im} z$. The operator $\tau_{AB}(z)$ denotes the "Coulomb-pole 
removed" transition operator that satisfies the equation 
\eq\label{coupledchannel}
& &\tau_{AB}(z)=v_{AB}+v_{A+}\hat g_{++}(z) \tau_{+B}(z)
+ \frac{1}{4}v_{A0} g_{00}(z) \tau_{0B}(z),\nonumber\\[2mm]
& &({\bf q}|\hat g_{++}(z)|{\bf p}) = ({\bf q}|g_{++}(z)|{\bf p})  
-\frac{\Psi_0({\bf q})\Psi_0({\bf p})}{(z-E_0)}\,.
\en
It remains to solve Eq.~(\ref{main}) in the dimensional regularization scheme. 
It is convenient to first reduce the two-channel equations 
(\ref{coupledchannel}) for the transition operators $\tau_{AB}(z)$ to the 
one-channel equation for $\tau_{++}(z)$ with an effective potential 
$w_{++}(z)$~\cite{Feshbach},
\eq\label{tau}
\tau_{++}(z)=w_{++}(z)+w_{++}(z)\hat g_{++}(z)\tau_{++}(z).
\en
The potential $w_{++}(z)$  includes the effects of the 
coupling to the $\pi^0\pi^0$ channel and has the form 
\eq\label{Wz}
&&({\bf q}|w_{++}(z)|{\bf p})=
(2\pi)^3\delta^3({\bf p}-{\bf q})\biggl(-\frac{{\bf p}^4}{4M_{\pi^+}^3}+
\cdots\biggr)\nonumber\\
&&\hspace*{.5cm}+w(z)+w_1(z){\bf p}^2+w_2(z){\bf q}^2
+w_3(z){\bf p}{\bf q}+\cdots\; .
\en
We find that the  corrections at order $\delta$ are generated entirely by the 
leading term in the connected part of the effective potential,
\eq\label{decaywidth}
& &\Gamma_{2\pi^0}=-\frac{\alpha^3 M_{\pi^+}^3}{4\pi}{\rm Im}w
\biggl(1+\frac{\alpha M_{\pi^+}^2}{4\pi}\xi
\, {\rm Re}w\biggr)+\cdots,\nonumber\\
& &\xi=2\ln\alpha-3+\Lambda+
\ln\frac{M_{\pi^+}^2}{\mu^2},\nonumber\\ 
& &\Lambda=(\mu^2)^{d-3}[(d-3)^{-1}-\Gamma'(1) - \ln4\pi],
\en
with $w= w(E_0)$. The divergent term proportional to $\Lambda$ 
stems  from a charged pion loop with one Coulomb photon exchange. 
It is removed by  the renormalization procedure in the scattering
sector, see below. Then, at the order of accuracy we are working here, 
the coefficient $w$ can be expressed in terms of the effective 
couplings $c_i$,
\eq\label{Re_Im}
{\rm Im}w&=&-\frac{M_{\pi^0}}{2\pi}\rho^{1/2}
\biggl( 1+\frac{5\rho}{8M_{\pi^0}^2}\biggr)(c_2-2\rho c_4)^2\nonumber\\
&\times&\biggl(1-\rho\,\frac{M_{\pi^0}^2c_3^2}{4\pi^2}\biggr),\,\,\,\,\, 
{\rm Re}w=-c_1,
\en
where
$\rho=2M_{\pi^0}(M_{\pi^+}-M_{\pi^0}-M_{\pi^+}\alpha^2/8)$. 

Next, we consider the matching procedure, which relates the effective 
couplings $c_i$  to the $\pi\pi$ scattering amplitude evaluated in the 
relativistic theory. To this end, we first calculate the $T$-matrix element  
for the process $\pi^+\pi^-\rightarrow\pi^0\pi^0$ at order $\delta$ in the 
non-relativistic theory, treating the Coulomb interaction as a perturbation. 
On the mass shell 
$z=(M_{\pi^+}^2+{\bf p}^2)^{1/2}=(M_{\pi^0}^2+{\bf q}^2)^{1/2}$, 
in the vicinity of the $\pi^+\pi^-$ threshold, this matrix element has the 
structure
\eq\label{structure}
\frac{B_1}{|{\bf p}|}+B_2\ln\frac{2|{\bf p}|}{M_{\pi^+}}
-\frac{1}{4M_{\pi^+}^2}{\rm Re}A^{+-00}_{\rm thr}\,+\cdots,
\en
where the ellipses stand for terms that vanish as ${\bf p}\rightarrow 0$. 
The singular contribution is generated by  Coulomb photon exchange.
The logarithmic term is absent at order $e^2p^2$ in which the calculations 
have been carried out in Ref.~\cite{Knecht} - it first appears at order 
$e^2p^4$~\cite{Roig}.

At order $\delta$, the quantity ${\rm Re}A^{+-00}_{\rm thr}$ is given by
\eq\label{A_thr}
& &\frac{1}{4M_{\pi^+}^2}\,{\rm Re}A^{+-00}_{\rm thr}=2c_2-4M_{\pi^0}^2
\kappa\biggl(c_4+\frac{c_2c_3^2}{8\pi^2}M_{\pi^0}^2\biggr) \nonumber\\
&+&\frac{\alpha M_{\pi^+}^2}{4\pi}\biggl(1-\Lambda- 
\ln\frac{M_{\pi^+}^2}{\mu^2}\biggr)c_1c_2.
\en
The divergent term $\Lambda$ may be absorbed in the coupling
$c_2$. This procedure eliminates at the same time the divergence in
the above expression  for the width. We may use the relation (\ref{A_thr}) 
to express the couplings $c_2$ and $c_4$ that occur in (\ref{Re_Im}) through 
${\rm Re}A^{+-00}_{\rm thr}$.

In order to also replace the couplings $c_1$ and $c_3$, we note that these are 
determined by matching  the amplitudes of the processes 
$\pi^+\pi^-\rightarrow\pi^+\pi^-$ and $\pi^0\pi^0\rightarrow\pi^0\pi^0$. 
It suffices to perform the calculation at $O(\delta^0)$, so that only the 
strong scattering lengths $a_0$ and $a_2$ appear in the matching condition: 
\eq\label{c1c3}
3M_{\pi^+}^2\, c_1&=&4\pi\, (2a_0+a_2) +o(\delta),\nonumber\\
3M_{\pi^+}^2\, c_3&=&2\pi\, (a_0+2a_2) +o(\delta).
\en
{}From Eqs.~(\ref{Re_Im}), (\ref{A_thr}) and (\ref{c1c3}) 
we find the following result for 
the quantities ${\cal A}$ and $K$ in  Eq.~(\ref{deser}), 
\eq\label{final}
{\cal A}&=&-\frac{3}{32\pi}\,{\rm Re}A^{+-00}_{\rm thr}+o(\delta),\\
K&=&\frac{\kappa}{9}\,(a_0+2a_2)^2
-\frac{2\alpha}{3}\,(\ln\alpha-1)\,(2a_0+a_2)+o(\delta).\nonumber
\en
Finally we note that, according to the matching condition, 
${\rm Re}A^{+-00}_{\rm thr}$ in this equation is {\it identical} to the 
corresponding quantity evaluated in the relativistic theory. According to 
(\ref{structure}), it  may  be  obtained by first calculating the relativistic
scattering amplitude for the process $\pi^+\pi^-\rightarrow\pi^0\pi^0$ at 
order $\delta$ near threshold, discarding  the singular pieces that behave 
like $|{\bf p}|^{-1}$ and $\ln 2|{\bf p}|/M_{\pi^+}$. The remainder, 
evaluated at the $\pi^+\pi^-$ threshold, equals ${\rm Re}A^{+-00}_{\rm thr}$ 
in Eq.~(\ref{final}). The calculation at order $e^2p^2$ has already been 
performed in \cite{Knecht} - the result is given in Eq.~(5.6) of this 
reference. 

The expressions  (\ref{final}) do not contain any divergences 
(in contrast to the findings of Ref.~\cite{Bunatian}), 
or any ambiguities related to the off-mass-shell
extrapolation of  Green functions in ChPT. All quantities entering ${\cal A}$ 
and $K$ are expressed via the on-mass-shell scattering amplitudes in ChPT 
through  relations that are valid at order $\delta$, and at all orders in the 
chiral expansion. Our final result for the width $\Gamma_{2\pi^0}$ contains 
all contributions up to and including terms of order $\delta^{9/2}$. In 
comparison, the width for the decay in two photons is of order 
$\delta^5$, namely, $\Gamma_{2\gamma}=\alpha^5M_{\pi^+}/4$.  

The structure of Eqs.~(\ref{deser}) and (\ref{final}) is the same as the one 
found previously~\cite{Sazdjian,Atom}. As for the factor $\cal A$, it 
obviously reproduces the result of~\cite{Sazdjian,Atom} at order 
$e^2p^2$ in the chiral expansion. However, the range of validity of 
 Eq.~(\ref{final}) is much wider: the derivation does not rely on the "local" 
approximation, and the result is valid to all orders in the chiral 
expansion. Comparing the correction factor $K$, we find that it
contains - in addition to the terms displayed in~\cite{Sazdjian,Atom}
-  (minor) contributions of order $\delta\, M_\pi^2 /F^2$. 
Finally, we note that the corrections evaluated  so far \cite{Kong,Holstein}
in the non-relativistic approach are seen to be a part of the total 
correction to the leading-order result. In Refs.~\cite{Sazdjian,Atom}  
these terms are referred to as "mass shift" corrections.

It is  our opinion that  Eqs.~(\ref{deser}) and (\ref{final}) finalize the 
attempts to calculate the width $\Gamma_{2\pi^0}$ at next-to-leading 
order, relegating the problem  to the evaluation  of the physical 
on-mass-shell scattering amplitude for the process 
$\pi^+\pi^-\rightarrow\pi^0\pi^0$ to any desired order 
in the chiral expansion.  A detailed derivation of the above results,
including a numerical analysis, will be provided in a separate publication
 \cite{Detail}.

In conclusion, we have evaluated the width $\Gamma_{2\pi^0}$ of the
$\pi^+\pi^-$ atom in its ground state at leading and next-to-leading
order in isospin breaking. The non-relativistic effective 
Lagrangian approach of Caswell and Lepage \cite{Lepage} appears to be
an extremely suitable tool for this purpose, that allows one 
to completely solve this problem. Its usefulness may be seen even more clearly
for the case of $p\pi^-$, $pK^-$, $d\pi^-$, $dK^-$ atoms, studied in ongoing 
or planned  experiments (PSI, KEK, DA$\Phi$NE), because this approach 
 trivializes the spin-dependent part of the problem. 

{\it Acknowledgments}. We are grateful to M. Beneke, H. Griesshammer, 
M. A. Ivanov, I. B. Khriplovich, T. Kinoshita, P. Labelle, G. P. Lepage, 
H. Leutwyler, K. Melnikov, P. Minkowski, L. L. Nemenov, M. Nio,
H. Sazdjian, J. Schacher,  J. Soto,  
J. Stern and O. V. Tarasov for useful discussions. 
V. E. L. thanks the University of Bern for hospitality.
This work was supported in part by the Swiss National Science
Foundation, and by TMR, BBW-Contract No. 97.0131  and  EC-Contract
No. ERBFMRX-CT980169 (EURODA$\Phi$NE).

\end{document}